\documentclass[a4paper,onecolumn,12pt,assc]{IEEEtranASSC}
\usepackage{cite}

\usepackage{graphicx}

\ifCLASSINFOpdf
\else
\fi
\usepackage[cmex10]{amsmath}
\usepackage{fixltx2e}

\usepackage{stfloats}

\usepackage[ampersand]{easylist}
\usepackage{subcaption}
\usepackage{cuted}

\newcommand{\dgr}{$^o $ }

\begin{document}
%
\title{Near real-time telecommand solutions for CubeSats: State of the art and applications to the SkyHopper mission}


\author{\IEEEauthorblockN{Robert Mearns,
Michele Trenti}

\IEEEauthorblockA{
School of Physics\\University of Melbourne\\Melbourne, Victoria}}


%


\maketitle

\begin{abstract}
Cubesats and similarly scaled nano-satellites present significant opportunities for hosting both scientific and commercial payloads for Earth sensing and astronomical observations, in particular in the area of rapid-response observations to external triggers. However, one limiting factor to full exploitation of the CubeSat potential in this area lies in the traditional approach of ground-spacecraft communications, which is based on infrequent contact via a limited network of ground stations. An alternative is to leverage existing commercial machine-to-machine orbital networks to transmit the triggers in near-real-time. Here, we present an analysis framework for calculating the likelihood of a time to first contact and the length of contact under minimum guaranteed conditions for these networks. The analysis is then extended to likely operational conditions in orbit, and the results of a comparative trade study of a number of orbital networks are presented, with an emphasis on the applicability to the SkyHopper Space Telescope CubeSat, a nanosatellite astronomical observatory currently undergoing preliminary design. It was found that near-real-time telecommands could be transmitted to SkyHopper within 10min with a likelihood of 62\% using the Globalstar network, or a likelihood of 74\% using the Iridium network, under predicted nominal operational conditions in orbit. Future networks currently under development could improve these figures to reach greater than 98\% coverage with a one second latency.

\end{abstract}



%
\IEEEpeerreviewmaketitle

\vspace{-0.4cm}
\section{Introduction}
\vspace{-0.4cm}
Traditional Telemetry and TeleCommand (TTC) schemes use either single ground stations, a network of ground stations, or government satellite relay networks such as NASA's Tracking and Data Relay Satellites (TDRS). However, for small missions built around NanoSatellites, the cost of utilising either a large network of ground stations, or TDRS is prohibitively expensive, while a single ground station can only provide regular but infrequent communication windows. Depending on orbit, these windows may occur only twice a day. For missions which require rapid communication for dissemination of observing targets these ground network contacts are insufficient. The opportunity then, presents itself to leverage existing commercial satellite relay networks to achieve the timeliness requirements of a mission. While both the GlobalStar and Iridium networks are currently being investigated on-board small-sats \cite{GEARRSReentry} \cite{TechEdSat}, no comprehensive study of all available networks has been published. To this end, five satellite phone networks are investigated in this paper to assess the potential for near-real-time trigger transmission: GlobalStar, Inmarsat, Iridium, Orbcomm and Thuraya. In addition a ground network of two polar stations, modelled on NASAs Near Earth Network (NEN), is included. The coverage at varying orbital altitudes, time to first contact and contact time, as well as the operational cost of using the networks are compared for each network. As a first application of the method, we analyse the network performance for communication with the SkyHopper Space Telescope Cubesat\footnote{http://skyhopper.space}.

\vspace{-0.4cm}
\section{SkyHopper Design Requirements}
\vspace{-0.4cm}
SkyHopper is a proposed 12U CubeSat equipped with a near IR telescope, currently funded for phase B study. One of its key science drivers is the follow-up of Gamma Ray Burst (GRB) triggers discovered by the Neil Gehrels Swift Observatory, with near-IR photometry, to identify candidate bursts at high redshift from the colours of their near-IR afterglow. In order to capitalise on the efficiency gain of prompt near-IR imaging that can be provided by space observations, triggers must be delivered to SkyHopper as rapidly as possible, since the afterglow rapidly fades as $\sim 1/t$. Given the limitations on telescope apertures in a CubeSat and therefore sensitivity limits, the scenario investigated here assumes that follow-up observations should commence within 10 minutes. To this end, any near-real-time communication system to be used must be capable of successful delivery of short bursts of data (~100 bytes) within such a time window. In addition, the availability of a message should be made evident to SkyHopper, either through a ring-alert system or storage of messages until SkyHopper can check for them and retrieve. Ideally, the near-real-time communication system would also be capable of bi-directional communication, though only half-duplex is required, so that rapid characterisations of the afterglow colours and resulting redshift estimates from on-board processing can be transmitted to the ground to activate spectroscopic follow-up observations of the most promising candidates at redshift $z>5$ (lookback time of 12.6 Gyr). The performance of various networks in achieving these goals is discussed for a number of typical nano-satellite LEO orbits, mentioned in section \ref{sec:minGuaranteedMethodology}, including that of SkyHopper, a 550km Sun synchronous orbit. Due to the low data rates and transient contact windows, this communication system would not be suited for telemetry or the majority of telecommand of SkyHopper, rather we limit our discussion to only the case of rapid triggers of transient observations.

\vspace{-0.4cm}
\section{State of the art Solutions}
\vspace{-0.4cm}
Of the 32 available communication relay networks, only five are suitable for nano-satellite use due to either constraints on their User Terminals (UTs), in terms of power consumption and size, type of provided service, network configuration or orbital configuration. These five networks are: GlobalStar, Iridium, Inmarsat, Orbcomm, and Thuraya. Each of these networks provide both simplex (UT $\rightarrow$ internet) and duplex (internet $\rightarrow$ UT $\rightarrow$ internet) machine-to-machine (M2M) services, or services which can be adapted for that purpose such as SMS. 
The following features of M2M which are of particular importance for rapid telecommand of smallsats in low earth orbit (LEO) are exhibited by each of these networks: short data scheduling times for timeliness of trigger/command data, store-then-forward capabilities to avoid trigger/command loss during periods of no contact and access for a large number of simultaneous users.
Of the five networks, GlobalStar is the only one for which dedicated small-sat components have been developed, primarily by Near Space Launch \cite{GEARRSReentry}, while COTS Iridium UTs have been modified to fly on a number of TechEdSat missions \cite{TechEdSat}, they have not been commercialised.
Inmarsat, Thuraya and Orbcomm networks also offer M2M capabilities designed for logistic asset tracking, but could be adapted for space-based assets. Unfortunately, other M2M networks such as Fleet Space or Myriota offer services to ground based assets only, or purely simplex services, allowing for real time telemetry, but not command of UTs.


A future alternative is Audacy, a promising network currently in development \cite{Audacy}. Their network would purportedly provide 100\% coverage to ground based assets and LEO satellites with data rates suitable for both TTC and mission data downlink, and a maximum transmission latency of one second. Their network is slated for completion by Q4 2019 \cite{Audacy}, and therefore has not been included in the following analyses.

\vspace{-0.4cm}
\section{Minimum Guaranteed Coverage Modelling}
\vspace{-0.4cm} \label{sec:minGuaranteedMethodology}
The time to first contact and the length of contact are the primary metrics in discriminating between the various networks. These metrics can be inferred from coverage figures, however, each of the five examined networks only provide coverage figures for a UT on the ground, not in orbit. As the altitude of a LEO satellite, and the UT which it carries, increases, the coverage figures and therefore the time to contact and length of contact times will decrease. The following methodology outlines how these figures are determined for candidate orbits.
The coverage provided by a network at discrete time intervals can be determined by the location of the UT within the network's pattern of beams, the link margin at this location, the ability for the network and UTs to compensate for the increased Doppler shift over the typical shift ground based UTs experience, and whether all these parameters remain with tolerable ranges for a sufficient period of time to complete a call set-up and communication. The call set-up collectively refers to the acquisition, synchronisation, authentication, and if necessary, location registration processes. In addition, the GlobalStar and Orbcomm networks, require simultaneous coverage of both the spacecraft UT and a network gateway on the ground, as they are bent-pipe networks.

	\begin{figure*}[!t]
		\centering
		\includegraphics[width=\textwidth]{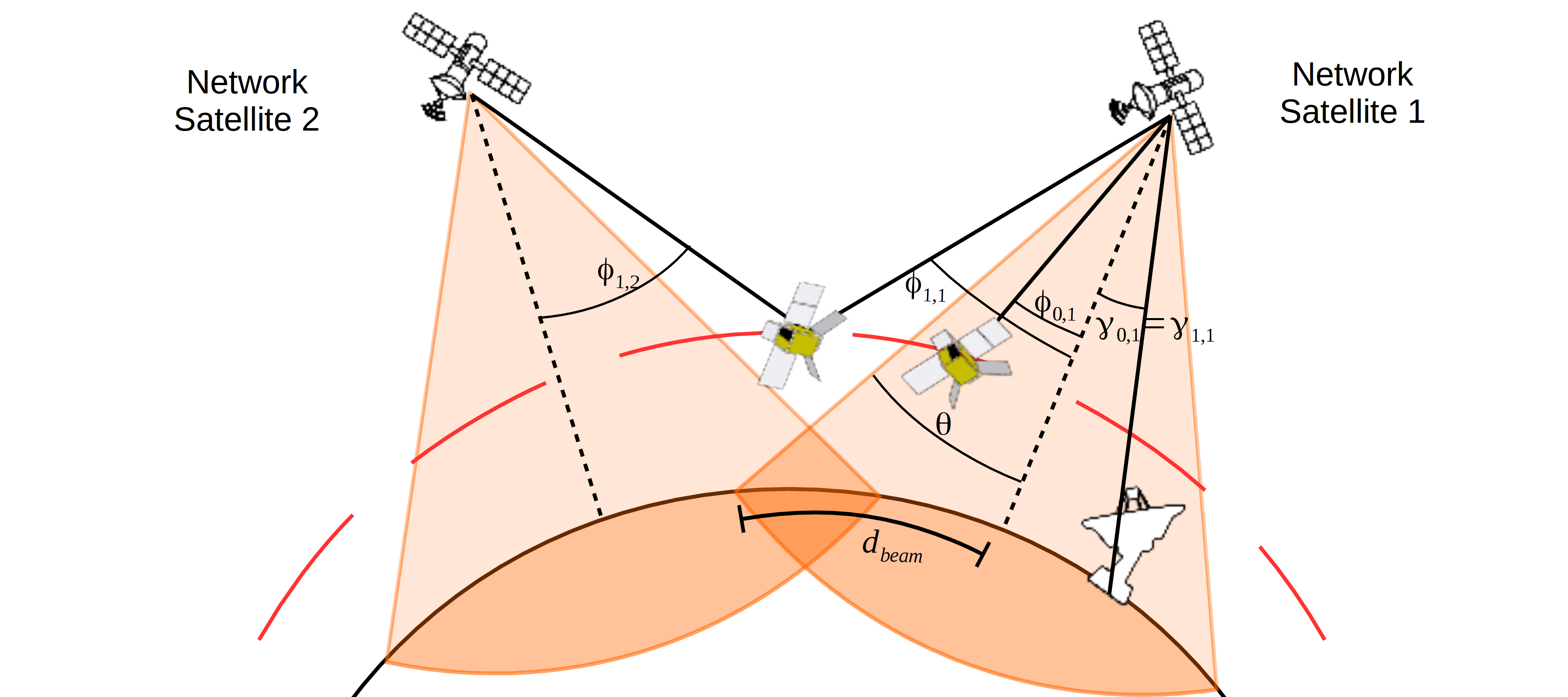}
		\caption{Constellation beam traversal by a nanosatellite carrying a network User Terminal (UT). At discrete time steps, $t=j$, the nanosatellite orbits along its trajectory (red dashed line). The angle $\phi_{i,j}$ measures the angle between the UT and the constellation satellite $i$ at timestep $j$. If $\phi_{j,i}<\theta$ the UT is within the constellation beam for satellite $j$. Additionally, for bent pipe networks, $\gamma_{j,i,k}$ measures the angle between satellite $i$'s nadir direction and gateway $k$.}
		\label{fig:beamDiagram}
	\end{figure*}

Figure \ref{fig:beamDiagram} outlines the situation of a nanosatellite carrying a UT traversing the constellation beams. As radiation patterns and link margins are not readily available for the satellites of all these networks, we begin by estimating the minimum guaranteed case, in which the constellation beams' beamwidth, $\theta$ is constant for all altitudes, and is given by the narrowest known possible beam-width $\theta_{min}$. 
It is known that on the ground, a UT at the edges of the ground spot enjoys sufficient link margin to allow for data transmission. In orbit, therefore, as long as the UT antenna is preferentially pointed, data transmission is feasible at least anywhere within the cone formed by the satellite and its ground spot. The apex angle of this cone, which can be determined from published ground spot sizes equals $\theta_{min}$. We perform a geometric analysis, referred to as the geometric visibility, over the course of a year, calculating $\theta$ for each time step and comparing it with the threshold values. In addition, for those bent-pipe networks, the angle between the network satellites' nadirs and the vectors to the network gateways, $\gamma$, are compared to the gateway's elevation threshold. An example of this is provided in figure \ref{fig:beamDiagram}. At $t=0$ both the satellite and gateway are visible to network satellite 1, as they both lie within the relevant beam. At $t=1$ the satellite has orbited past the extents of the beam and is no longer visible to either network satellite, however, the gateway remains visible to network satellite 1.
This analysis is modified slightly in the case of the Iridium constellation. Here network beams switch off when over the poles to minimise interference while maintaining coverage on the ground. This behaviour is achieved by switching the beams off on satellites in orbital planes 2,4 and 6 when the satellite is below -70 \dgr latitude and on satellites in orbital planes 1,3 and 5 when the satellite is above 70 \dgr latitude. The Iridium orbital planes are indexed at 1 anticlockwise from the orbital seam when viewed from the north pole \cite{IridiumFCCStatement}. This behaviour is included in our analysis by masking those satellites which are active at each time step.

The ability for the network satellites to compensate for Doppler shift is modelled by the Doppler visibility. It is calculated from the satellite's rate of change of range to each network satellite. Although the combined orbital velocity of the LEO satellite and network satellites occasionally result in a Doppler shift greater than that which is tolerable by the UT, these periods do not occur during those times where the UT is within the network satellite's beam. This behaviour is as a result of the small region of beam traversal being centred on the constellation satellite's nadir, resulting in a small relative rate of change of range, and therefore small Doppler shift during these periods.
The geometric and Doppler visibilities, and if necessary the gateway visibility, are combined via a logical AND operation for each time step and constellation satellite, resulting in the total UT visibility at each time step. A mathematical expression of this analysis is provided here. 

For time steps $t\in \{1,2,... ,j\}$, network satellites ($i$) and network gateways ($k$), the total visibility is given by:
\vspace{-0cm}
\begin{align*} 
v &=vis_{geom} \wedge vis_{gtwy} \wedge vis_{dopp}\\
v &=\left[\begin{smallmatrix}f(1,1) & \dots & f(1,j) \\
	\vdots & \ddots & \vdots \\
	f(i,1) & \dots & f(i,j) \\
	\end{smallmatrix}\right] \wedge
	\left[\begin{smallmatrix}g(1,1) & \dots & g(1,j) \\
	\vdots & \ddots & \vdots \\
	g(i,1) & \dots & g(i,j) \\
	\end{smallmatrix}\right] \wedge
	\left[\begin{smallmatrix}d(1,1) & \dots & d(1,j) \\
	\vdots & \ddots & \vdots \\
	d(i,1) & \dots & d(i,j) \\
	\end{smallmatrix}\right]\\
\quad &=\begin{bmatrix}
		v_{(1,1)} & \dots & v_{(1,j)} \\
		\vdots & \ddots & \vdots \\
		v_{(i,1)} & \dots & v_{(i,j)} \\
		\end{bmatrix} 
\end{align*}
\begin{center}
where $\begin{cases}
	f(i,j)=[\phi_{ij}<\theta]\\
	g(i,j)=\sum_{k}{[\gamma_{ijk}<\theta]}\\
	d(i,j)=[\frac{\dot{r_{ij}-r_{i,sh}}}{\lambda_j}<\Delta \nu_{max}]\\
	\end{cases}$, from figure \ref{fig:beamDiagram}

and	$[x<y] = $
		$\begin{cases}
			1, x<y\\
			0, x>=y           
		\end{cases} $\\
\end{center}
Accounting for the call set-up time is achieved by first performing a summation across each of the network satellites for each time step, and then finding sequences of contact within this $[1 \times i]$ vector, which are longer than the time required for call set-up. Sequences which are shorter than the call set-up time are discarded by setting them equal to zero.
\noindent The parameters for each network in the minimum guaranteed case are listed in table \ref{tab:minimumParam}.

	\begin{table}[!t]
	\small
	\renewcommand{\arraystretch}{1.3}
	\caption{Minimum Guaranteed Coverage Parameters}
	\label{tab:minimumParam}
	\centering
	\begin{tabular}{lccc}
	\hline
	Network 	& \parbox[bc]{1.4cm}{Half Beam-width \dgr}&\parbox[bc]{1.0cm}{Call set-up time (s)}& \parbox[bc]{1.0cm}{Max tol. Doppler (kHz)} \\
	\hline
	Globalstar 	& 54 	& 28	&46.5\\
	Iridium 	& 62.9 	& 7		&37.5\\
	Inmarsat 	& 8.6	& 15	&37.5\\
	Orbcomm 	& 62	& 20	&25.3\\
	Thuraya 	& 8.6	& 15	&37.5\\
	Ground Station 	& 80	& 5 & 41.0\\
	\hline
	\end{tabular}
	\end{table}

Once the total visibility has been calculated for the entire year, the total visibility is sampled using a Monte-Carlo simulation consisting of 50,000 uniformly distributed events. For each sample point, both the time to first contact and the length of the contact are recorded. In this way, the probability of being able to contact a UT (in a particular orbit) within a certain time is found.

Both the visibility analysis and the Monte-Carlo simulation were performed for a number of candidate orbits typical of nano-satellite launches. Each candidate orbit was selected from a range of altitudes (400-700km), varying RAAN of Sun-synchronous orbits (dawn-dusk and 1030LTDN), varying inclination (Sun-synchronous, International Space Station) In all cases the argument of perigee was also varied to test the effect of orbital phase lag on the the overall coverage. All network satellite positional data was generated using historical TLEs and propagating these into the future, while nano-satellite orbits were propagated from the candidate orbital elements.

	\subsection{Minimum Guaranteed Coverage Results} \label{sec:minGuaranteedResults}
	\vspace{-0.4cm}
	\begin{table}[!t]
	\small
	\renewcommand{\arraystretch}{1.3}
	\caption{Minimum Guaranteed Coverage Results for a 550km Sun-Synchronous Dawn-Dusk Orbit}
	\label{tab:minGuaranteedResults}
	\centering
	\begin{tabular}{lcccc}
	\hline
	{} & \multicolumn{4}{c}{Time (secs) required for probability of contact}  		\\
	\hline
	Network 		& 50 \% 		& 68 \% 		&  95 \% 	& 99.7 \%			\\
	\hline
	Globalstar 		& 915			& 1676 			& 4328		& 7169 				\\
	Iridium 		& 838 			& 1422 			& 2995 		& 6498 				\\
	Inmarsat 		& 0				& 524  			& 1440 		& 3629 				\\
	Orbcomm 		& 2893			& 4926 			& 12445 	& 22503 			\\
	Thuraya 		& 2855			& 4285 			& 15295 	& 20932 			\\
	Ground Station 	& 1675 			& 2523 			& 4754 		& 5195 				\\
	\hline
	{} 				& {}			& {} 			& {} 		& {} 			  	\\
	\hline
	{} & \multicolumn{4}{c}{Probability of contact within certain time} \\
	\hline
	{} 				& \multicolumn{2}{c}{Immediately}   &1 min 	& 10 min 	\\
	\hline
	Globalstar 		& \multicolumn{2}{c}{10.3 \%} & 13.8 \% 	& 39.2 \% 	\\
	Iridium 		& \multicolumn{2}{c}{7.8 \%} & 13.4 \% 	& 42.0 \% 	\\
	Inmarsat 		& \multicolumn{2}{c}{50.0 \%} & 52.1 \% 	& 70.4 \% 	\\
	Orbcomm 		& \multicolumn{2}{c}{1.4 \%} &  2.8 \% 	& 13.8 \% 	\\
	Thuraya 		& \multicolumn{2}{c}{13.7 \%} & 14.5 \% 	& 21.4 \% 	\\
	Ground Station 	& \multicolumn{2}{c}{9.5 \%} & 11.0 \% 	& 24.0 \% 	\\
	\hline
	\end{tabular}
	\end{table}

\begin{figure*}[!t]
		\centering
		\includegraphics[width=\textwidth]{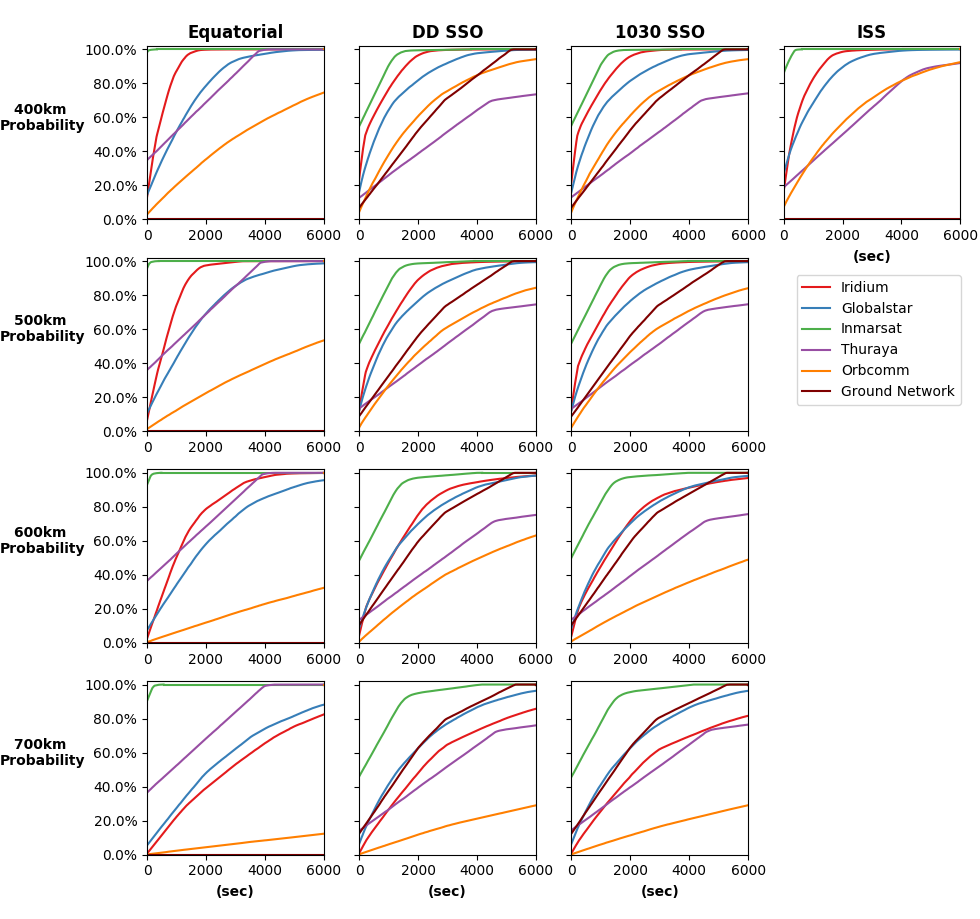}
		\caption{Cumulative probability of a particular time to first contact under minimum guaranteed coverage conditions for all investigated networks.}
		\label{fig:minCumTimeToFirstContact}
	\end{figure*}

\begin{figure*}[!t]
		\centering
		\includegraphics[width=\textwidth]{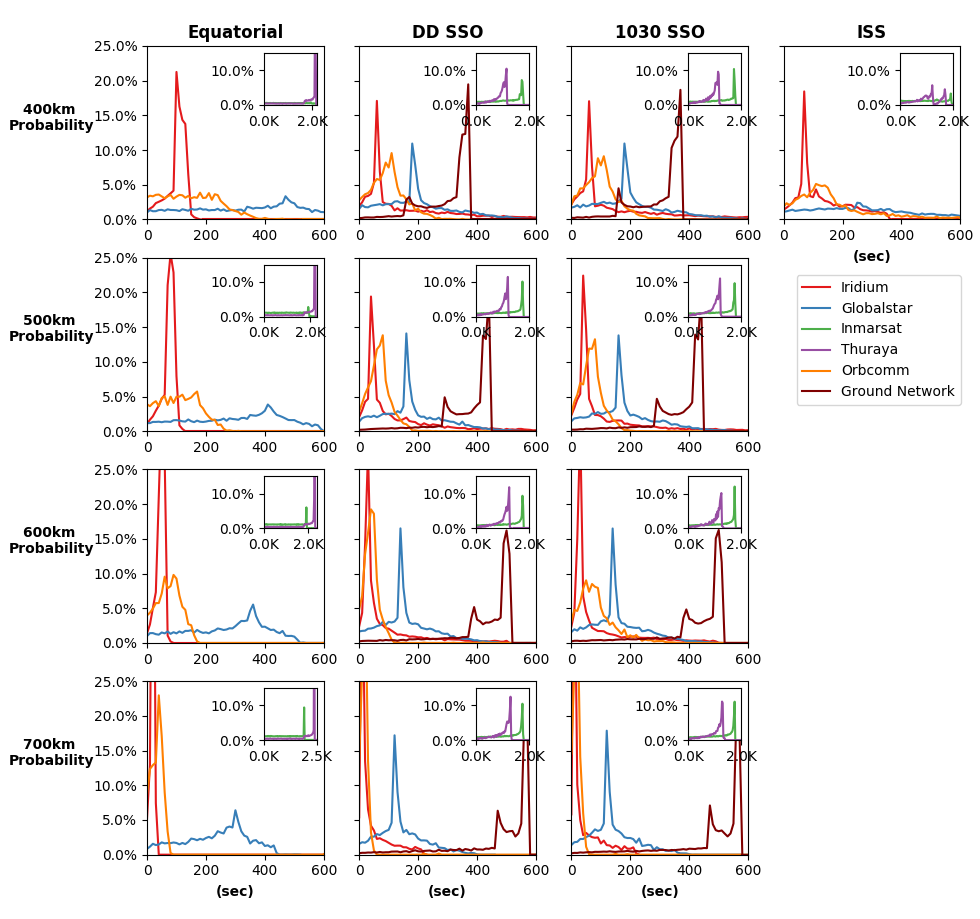}
		\caption{Probability of the first contact time under minimum guaranteed coverage conditions for all investigated networks.}
		\label{fig:minFirstContactTime}
	\end{figure*}

	This section outlines the results of the minimum guaranteed coverage analysis described in section \ref{sec:minGuaranteedMethodology}
	The plots in figure \ref{fig:minCumTimeToFirstContact} illustrate the probability of a randomly occurring trigger successfully being transmitted to our UT after a certain period of time has elapsed, while figure \ref{fig:minFirstContactTime} shows the distribution of contact times for the first contact with each network. The width of the distribution peak gives an indication of how widely spread the range of contact times is.
	All orbital networks experience a reduction in time to first contact and a decrease in contact time as the altitude of the UT increases, since the satellite is traversing narrower sections of the beam cone as altitude increases. The ground network experiences the opposite trend, due to its inverted beam cone, increasing from the modal contact time of 380 seconds to 575 seconds at an altitude of 700km.
	As we would expect, the probability of time to first contact for the Iridium Network is largely independent of inclination; at an altitude of 400km, the probability of immediate contact is within 20\% to 30\% across the candidate inclinations. This behaviour is likely due to the intra-satellite spacing within their orbital planes, and the spacing of the orbital planes themselves being uniform and constant. In comparison, the Globalstar and Orbcomm networks, since they have orbital paths with a relatively low inclination, have slight variations in their time to first contact probabilities with inclination.
	For all orbital networks, candidate orbits which fall between $\pm$ 70 \dgr latitude, experience a slightly lower time to first contact, and greater contact time. This is primarily due to these networks being designed to operate for UTs on the ground between these latitudes. For example, the GlobalStar satellites' orbits have an inclination of 52 \dgr; the equatorial and ISS candidate orbits in this case reach a cumulative probability of time to first contact of 99.7\% within approximately 80 minutes, while the polar orbits take over 100 minutes to reach this probability. This is due to the fact that while a UT in a polar orbit is between $\pm 70 $ \dgr $\rightarrow 90$ \dgr, it will not experience coverage with these networks, reducing the total coverage as compared to a similar orbit but at a reduced inclination. This effect impacts the GlobalStar, Inmarsat, Thuraya and Orbcomm networks. The Iridium network suffers from a similar reduction for polar orbits, however, this is due to the shut down of satellites while in this region. While this does not have an effect on UTs on the ground as total coverage there is not reduced due to overlaps in the beam, at higher altitudes the reduced number of active beams has a noticeable effect. In contrast, due to the polar placement of our Ground Network, there is zero contact for non polar orbits.
	The polar location of the Ground Network unfortunately, can not be used to supplement any of the orbital networks suffering from reduced coverage over the Arctic regions, despite the Ground Network coverage being high in these regions. Since the NEN requires at least a several day lead time on data packet preparation, it is unsuitable for randomly occurring triggers.

\vspace{-0.4cm}	
\section{Likely Operational Coverage Modelling}
\vspace{-0.4cm}
Although the minimum guaranteed coverage, outlined in section \ref{sec:minGuaranteedMethodology} is useful for discriminating between the different constellations, it overestimates the time to contact due to several conservative estimates made in the analysis methodology. This section outlines modifications to the minimum guaranteed coverage modelling to more accurately represent the likely operational case.
Of the five investigated constellations, Inmarsat, Iridium and Globalstar present the highest coverage potential and lowest time to contact for those candidate orbits likely to be used for the SkyHopper mission. However, as both Iridium and Globalstar UTs have successfully flown on cubesats before, and custom Inmarsat applications require lengthy and extremely stringent approval processes, only the Iridium and Globalstar networks are examined further. 
\vspace{0.8cm}
The call set-up times of both the Iridium and Globalstar constellations are overestimated in section \ref{sec:minGuaranteedMethodology}, resulting in shorter contact periods being unnecessarily discarded when calculating the minimum guaranteed case, modifications are addressed in section \ref{sec:callSetUpModifications}. Additionally, the minimum guaranteed coverage analysis underestimates the spatial coverage of each constellation beam, due to assuming a constant beam width for all altitudes, a more accurate representation of the Iridium and Globalstar beams is addressed in section \ref{sec:likelyBeamWidth}.

	\vspace{-0.6cm}
	\subsection{Call set-up time} \label{sec:callSetUpModifications}
\begin{figure*}[!t]
			\centering
			\includegraphics[width=0.5\textwidth]{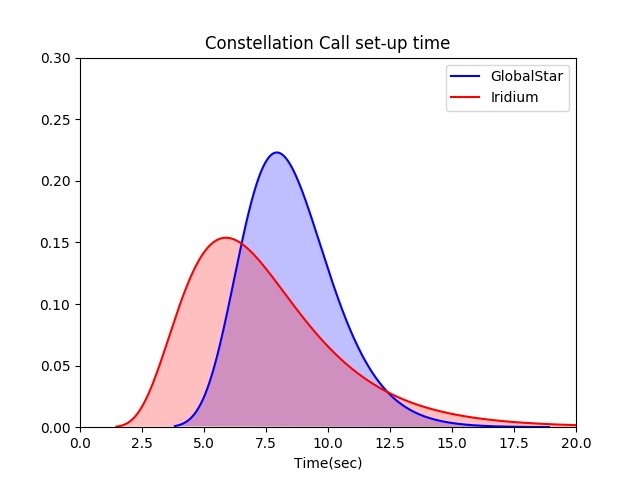}
			\caption{Call set-up time probability distributions for M2M type messages within the GlobalStar and Iridium Networks, \cite{GEARRSReentry},\cite{IridiumLatency}.}
			\label{fig:callSetupTimePdf}
		\end{figure*}

	\vspace{-0.4cm}
	While the original estimates of the call set-up time used the mean theoretical set-up time for each network, this is not the case under operational conditions.	The set-up times for a M2M type message within the Iridium and GlobalStar networks instead depend on a search through the parameter spaces governing the channel reuse schemes for each network. Consequently, the set-up times are distributed according to the probability density functions shown in figure \ref{fig:callSetupTimePdf}, constructed from data in \cite{IridiumLatency} and \cite{GEARRSReentry}. As such, the most probable time for call set-up is shorter than the mean time used for the minimum guaranteed contact analysis.
	Iridium set-up times include the time taken to transmit a 100 byte (at 1.2kbps) Mobile Originated (MO) message required to check the mailbox for waiting Mobile Terminated (MT) messages. This is the use case for the SkyHopper observatory due to contact blackouts with the Iridium network. However, for lower orbits which can maintain contact, the ring-alert functionality would be sufficient, and as such the call set-up times would be fractionally shorter.
	For each event of the Monte-Carlo simulation, a new set-up time is generated from the relevant distribution.

	\vspace{-0.6cm}
	\subsection{Likely beam-width modelling} \label{sec:likelyBeamWidth}
\begin{figure*}
		\begin{subfigure}{0.5\linewidth}
			\centering
			\includegraphics[width=\linewidth]{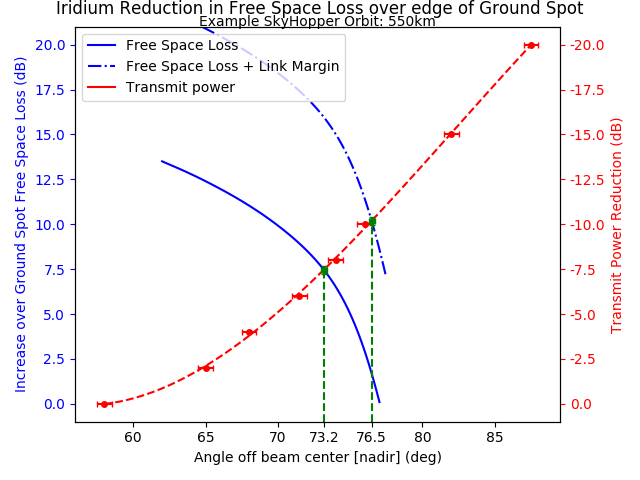}
			\caption{Iridium}
			\label{sfig:angleIR}
		\end{subfigure}\hfill
		\begin{subfigure}{0.49\linewidth}
			\centering
			\includegraphics[width=\linewidth]{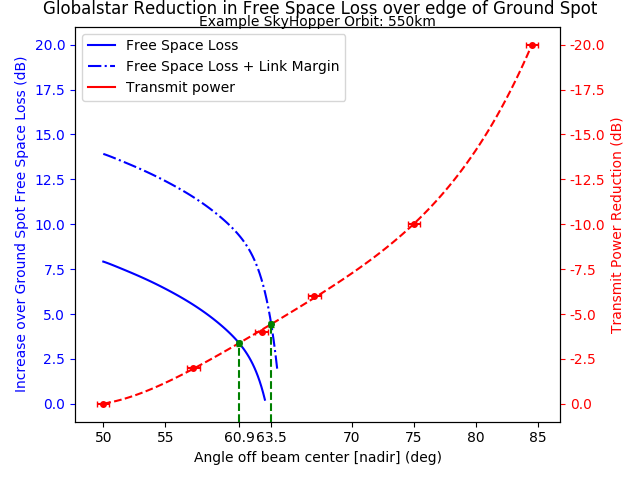}
			\caption{Globalstar}
			\label{sfig:angleGB}
		\end{subfigure}\hfill
		\caption{Reduction in Free Space Loss over edge of Ground Spot for UT altitude of 550km}
		\label{fig:ReducFSPL}
	\end{figure*}

\begin{figure*}
		\begin{subfigure}{0.49\linewidth}
			\centering
			\includegraphics[width=\linewidth]{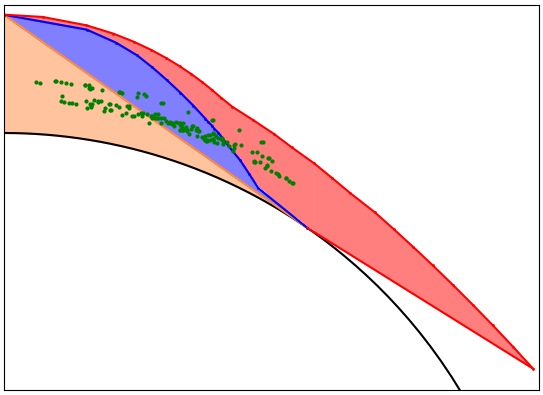}
			\caption{}
			\label{sfig:GlobalstarWideBeam}
		\end{subfigure}\hfill
		\begin{subfigure}{0.49\linewidth}
			\centering
			\includegraphics[width=\linewidth]{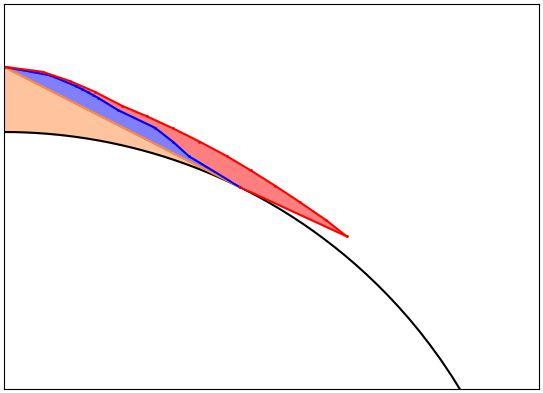}
			\caption{}
			\label{sfig:IridiumWideBeam}
		\end{subfigure}\hfill
		\caption{Likely operational beam width for constellation satellites. The minimum guaranteed beam width is given by the orange region, the beam width increase from the compensation by free-space-path-loss reduction with link margin is given by blue region. The maximum possible increase in beam-width due to compensation, and including consumption of link margin is given by the red region. \emph{a)}: Globalstar Network. Green dots represent GEARRS 2 angle from constellation satellite during contact. 11\% of GEARRS2 contacts occurred outside the nominal likely beam width, but within the maximum likely beam width. \emph{b)}: Iridium Network}
		\label{fig:wideBeam}
	\end{figure*}

	\vspace{-0.4cm}
	The minimum guaranteed coverage case estimated the constellation beam widths, hereafter referred to as the minimum beam widths, using the ground spot size and knowledge that the link margin is sufficient to communicate at the edge of the ground spot. However, as the range between the UT carrying nanosatellite and the constellation satellite will be at a fraction of the slant rage of the constellation satellite to the edge of the ground-spot, the reduction in free space path loss can compensate for the reduction in transmit power further off axis than the minimum beam width, thereby maintaining the same link margin as for a subscriber unit at the edge of the ground-spot. Consequently, the likely operational nominal beam-width will be wider than that of the minimum beam width, and a function of the UT altitude.
	Additionally, if the available link margin to maintain the lowest data rate within a network is consumed by the free space loss, a UT could communicate even further off the beam-axis than the nominal beam width. This maximum beam width represents the widest possible beam width at each altitude while maintaining operation of the network.
	In order to calculate both the nominal and maximum likely operational beam widths, the antenna radiation pattern for the network satellites is used. The antenna radiation patterns for the Iridium and Globalstar satellites can be found in the FCC licence to operate submissions, \cite{IridiumFCCStatement}, \cite{GlobalstarFCC}. The reduction in free space loss for any given altitude is found from the triangle formed between the centre of the earth, a constellation satellite and nanosatellite. The curve obtained by solving for the constellation satellite-nanosatellite range as a function of angle off beam axis, for a given nanosatellite altitude, can be converted to a decibel curve for each constellation. The reduction in free space loss curve and the reduction in antenna transmit radiation pattern curve can be plotted against each other, and the intersection of these two curves will yield the nominal likely operational beam-width for that altitude. Including the budgeted link margin into the reduction in free space loss lets us solve for the maximum likely operational beam-width for that altitude. The Iridium and Globalstar curves for a nanosatellite orbit with altitude of 550km are given in figures \ref{sfig:angleIR} and \ref{sfig:angleGB}, while the likely operational beam widths for both Iridium and Globalstar constellations for the range of candidate orbits likely to be used by the SkyHopper mission are given in table \ref{tab:likelyOperationalBeamWidth}.
	To validate this beam width estimation, positional data from the GEARRS 2 satellite, kindly provided courtesy of Near Space Launch, during its $350~\mathrm{km} \times 700~\mathrm{km}$ orbit was converted to the angle off axis from the Globalstar satellite it was communicating with. This was then cross-referenced with the time at which simplex communication via the Globalstar network was known to have occurred. These points are shown in figure \ref{sfig:GlobalstarWideBeam}, in all cases, these points lie within the maximum likely operational beam.
	
	\begin{table}[!t]
	\small
	\renewcommand{\arraystretch}{1.3}
	\caption{Likely Operational Beam width for SkyHopper altitude Range}
	\label{tab:likelyOperationalBeamWidth}
	\centering
	\begin{tabular}{ccc}
	\hline
	Altitude (km)	& Iridium beam-width (\dgr) & Globalstar beam-width (\dgr)\\
	\hline
	500 	& $71.7_{-0.2}^{+2.9}$ & $60.2_{-0.1}^{+2.6}$	\\
	525 	& $72.4_{-0.2}^{+2.8}$ & $60.5_{-0.2}^{+2.7}$	\\
	550 	& $73.2_{-0.2}^{+3.3}$ & $60.9_{-0.2}^{+2.6}$	\\
	575 	& $74.0_{-0.2}^{+2.7}$ & $61.3_{-0.2}^{+2.8}$	\\
	600 	& $74.9_{-0.2}^{+2.7}$ & $61.7_{-0.1}^{+2.5}$	\\
	\hline
	\end{tabular}
	\end{table}

	\vspace{-0.6cm}
	\subsection{Likely Operational Coverage Results}
\begin{figure*}[!t]
		\centering
		\includegraphics[width=\textwidth]{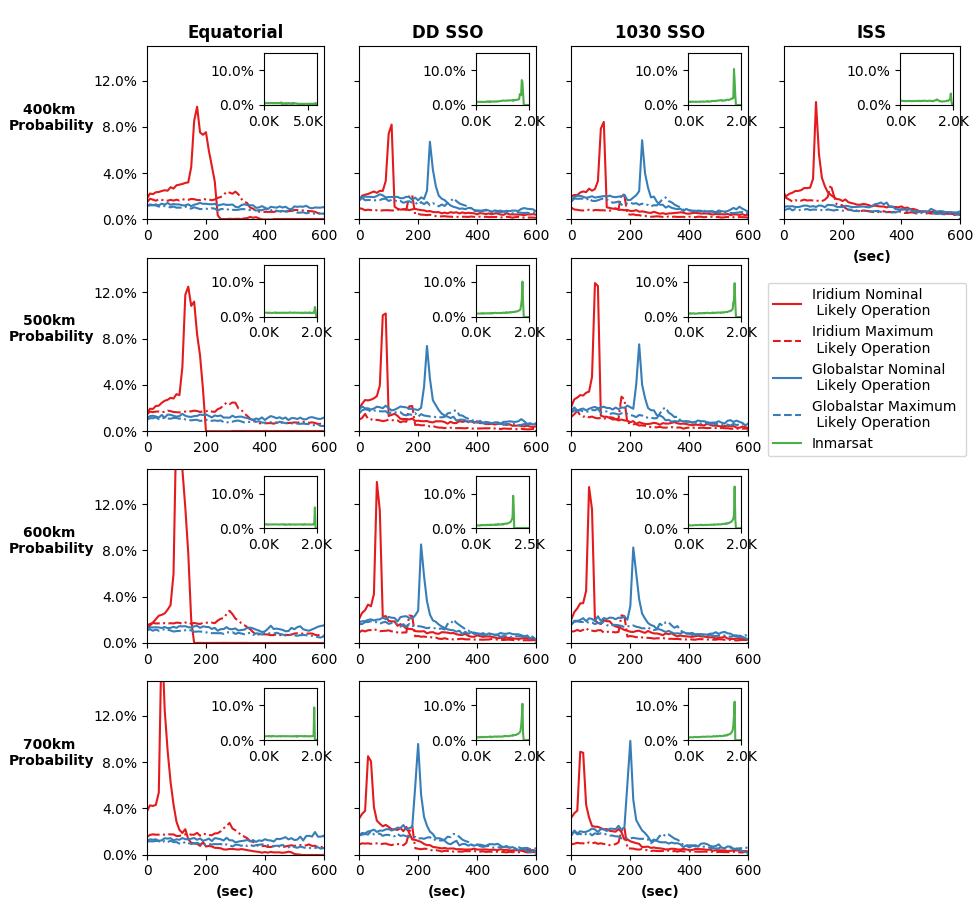}
		\caption{Probability of the first contact time under Maximum and Nominal operational coverage conditions for the Globalstar and Iridium Networks. The minimal guaranteed operational case for the Inmarsat network is provided in green as a reference.}
		\label{fig:likelyFirstContactTime}
	\end{figure*}

	\vspace{-0.4cm}
	Similar to section \ref{sec:minGuaranteedResults} this section outlines the results of the likely operational coverage analysis, subject to the modifications described in section \ref{sec:callSetUpModifications} and \ref{sec:likelyBeamWidth}.
	The plots in figure \ref{fig:minCumTimeToFirstContact} and \ref{fig:likelyFirstContactTime} illustrate the probability of a particular time to first contact and the length of that contact for each network, respectively.

	Both the nominal and maximum likely operational cases experience a dramatic reduction in the time to first contact with all candidate orbits reaching a 99.7\% likelihood of contact within 85 minutes, as compared to the minimum guaranteed coverage, in which only the lowest altitude orbits had reached 99.7\% likelihood within 85 minutes for the Globalstar and Iridium networks. 
	The time to first contact under only the maximum likely operational case operating within the Iridium network, will be less than 7 minutes in all cases, and operating within the Globalstar network will be less than an hour in all cases.
	For both networks the nominal likely operational case experiences a delay in the time to first contact when compared to the maximum likely case, this effect increases with increasing altitude. It is of interest to note that the 700km candidate altitude orbits for the Iridium constellation appear to contradict this trend. However, as the constellation satellite altitude is only 780km, the nominal and maximum beam widths suddenly converge at this altitude, as can be seen in figure \ref{sfig:IridiumWideBeam}.
	The distribution of contact times for each operational case and network becomes narrower with increasing altitude, similar to the trend in the minimum guaranteed case. However, due to the extremely large beam widths of the maximum operational cases, the distribution of contact times is very broad.

	For polar orbits similar to that of SkyHopper, both the Iridium and Globalstar networks perform similarly to the minimum guaranteed coverage case of the Inmarsat network. Although this is a comparison of networks under different operational conditions, it is unlikely that in the case of the Inmarsat network, coverage would increase dramatically. The Inmarsat network satellites operate under extremely stringent power requirements due to the massive free space path loss experienced by geostationary satellites. Consequently, their antennas are extremely directional. This, combined with the extreme distance to the satellites is likely to result in very little potential beam width increase due to reduction in free space path losses compensating for off-axis transmit power reductions, as described in section \ref{sec:likelyBeamWidth}. Finally, as there are only three effective Inmarsat satellites, any potential increase in beam width, though small, would only be experienced infrequently. Consequently, we feel it is valid to compare the minimum guaranteed coverage case of the Inmarsat network with the nominal likely operational cases of LEO networks.

	\begin{figure*}[!t]
		\centering
		\includegraphics[width=\textwidth]{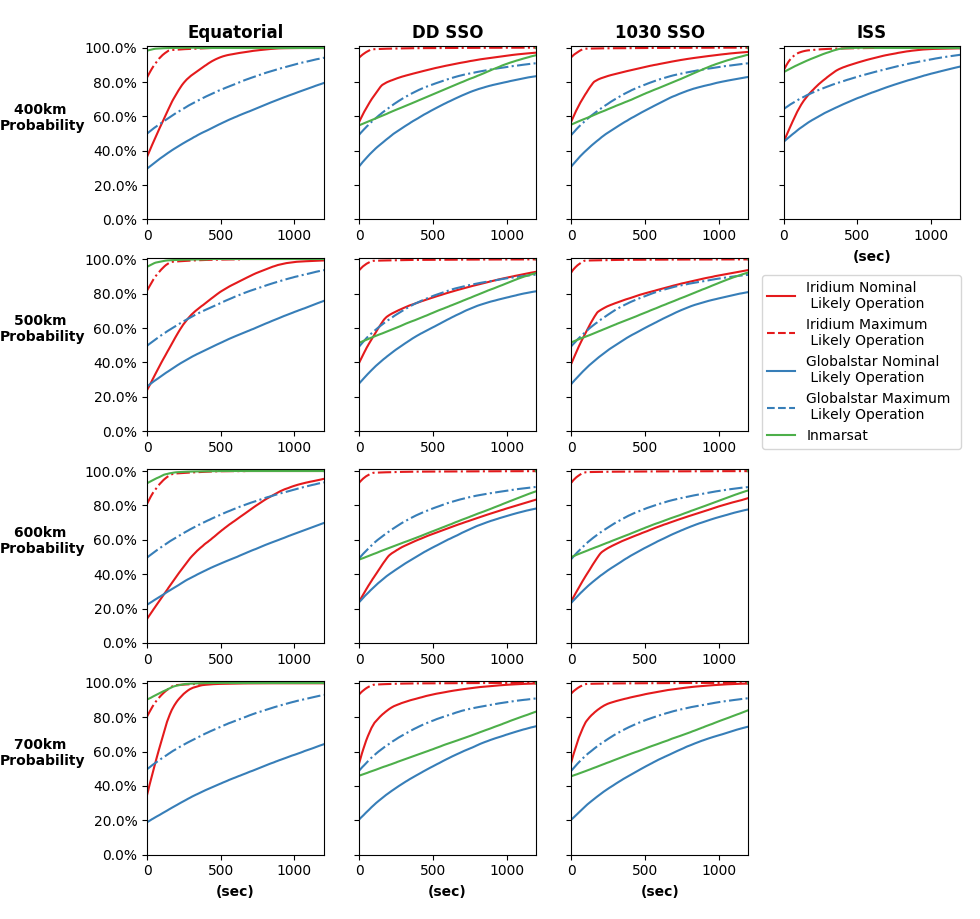}
		\caption{Cumulative probability of a particular time to first contact under Maximum and Nominal operational coverage conditions for the Globalstar and Iridium Networks. The minimal guaranteed operational case for the Inmarsat network is provided in green as a reference.}
		\label{fig:likelyCumTimeToFirstContact}
	\end{figure*}

	\begin{table}[!t]
	\small
	\renewcommand{\arraystretch}{1.3}
	\vspace{0.2cm}
	\caption{Nominal and Maximum Likely Operational Coverage Results for a 550km Sun-Synchronous Dawn-Dusk Orbit}
	\label{tab:nomMaxOperationalResults}
	\centering
	\begin{tabular}{lcccc}
	\hline
	{} & \multicolumn{4}{c}{Time (secs) required for probability of contact}  		\\
	\hline
	Network 		& 50 \% 		& 68 \% 		&  95 \% 	& 99.7 \%			\\
	\hline
	Globalstar Nominal 		& 343			& 748 			& 2857		& 4059 				\\
	Globalstar Maximum 		& 10			& 262 			& 1753		& 3296 				\\
	Iridium Nominal	& 121 			& 396 			& 1522 		& 2394 				\\
	Iridium Maximum	& 0 			& 0 			& 16 		& 366 				\\
	\hline
	{} 				& {}			& {} 			& {} 		& {} 			  	\\
	\hline
	{} & \multicolumn{4}{c}{Probability of contact within certain time} \\
	\hline
	{} 				& \multicolumn{2}{c}{Immediately}   &1 min 	& 10 min 	\\
	\hline
	Globalstar Nominal	& \multicolumn{2}{c}{25.4 \%} & 30.9 \% 	& 62.1 \% 	\\
	Globalstar Maximum	& \multicolumn{2}{c}{49.0 \%} & 54.5 \% 	& 81.2 \% 	\\
	Iridium Nominal		& \multicolumn{2}{c}{31.8 \%} & 41.4 \% 	& 74.2 \% 	\\
	Iridium Maximum		& \multicolumn{2}{c}{93.4 \%} & 97.8 \% 	& 99.8 \% 	\\
	\hline
	\end{tabular}
	\end{table}

\vspace{-0.4cm}
\section{Technical Constraints}
\vspace{-0.4cm}
It should be noted that in order to be permitted to use a network's UTs in orbit, limits may be imposed by the network on the transmission power of orbiting UTs so as to not overwhelm the network's receivers. However, this could be achieved either through gain controls on the UT itself, or an external attenuator between the UT and antenna.

\vspace{-0.4cm}
\section{Conclusion}
\vspace{-0.4cm}
While nano-satellites offer significant potential for Earth and astronomical observation missions, certain mission types require near-real-time telecommand. In this paper, we presented a framework for analysing the probability distribution of contact time lag between a network and a satellite, depending on the network architecture and satellite orbit. A first application to the SkyHopper mission shows that among commercially available solutions, the Iridium and Globalstar networks provide the highest likelihood of delays in commanding, while the geostationary networks, Inmarsat and Thuraya offer extremely long contact windows. For typical nanosatellite orbits, commands can be guaranteed to be delivered to the spacecraft within 1 hour 70\% of the time, while under predicted nominal conditions, commands can be delivered within just a few minutes 70\% of the time, using the Iridium network. For the SkyHopper mission, either of the Iridium or Globalstar networks is estimated to provide delivery of commands with less than 10 min lag at greater than 62\% confidence. This performance can be improved by a combination of the two networks, or by the use of future solutions for near-real time communications such as the Audacy network which should be available by early 2020 \cite{Audacy}.

\vspace{-0.6cm}
\section*{Acknowledgements}
\vspace{-0.4cm}
The authors would like to thank the Laby Foundation for supporting this work, Near Space Launch for providing GEARRS 2 performance characteristics, and the two anonymous reviewers for useful comments and suggestions.

\bibliographystyle{IEEEtran}
\vspace{-0.4cm}
\bibliography{2017_mearns_asrc_paper}

\begin{thebibliography}{1}
\providecommand{\url}[1]{#1}
\csname url@samestyle\endcsname
\providecommand{\newblock}{\relax}
\providecommand{\bibinfo}[2]{#2}
\providecommand{\BIBentrySTDinterwordspacing}{\spaceskip=0pt\relax}
\providecommand{\BIBentryALTinterwordstretchfactor}{4}
\providecommand{\BIBentryALTinterwordspacing}{\spaceskip=\fontdimen2\font plus
\BIBentryALTinterwordstretchfactor\fontdimen3\font minus
  \fontdimen4\font\relax}
\providecommand{\BIBforeignlanguage}[2]{{%
\expandafter\ifx\csname l@#1\endcsname\relax
\typeout{** WARNING: IEEEtran.bst: No hyphenation pattern has been}%
\typeout{** loaded for the language `#1'. Using the pattern for}%
\typeout{** the default language instead.}%
\else
\language=\csname l@#1\endcsname
\fi
#2}}
\providecommand{\BIBdecl}{\relax}
\BIBdecl

\bibitem{GEARRSReentry}
H.~Voss, J.~Dailey, M.~Orvis, A.~White, and S.~Brandle, ``Globalstar link: From
  reentry altitude and beyond,'' \emph{Proceedings of the 30th AIAA/USU
  Conference On Small Satellites}, no. SSC16-WK-11, 2016.

\bibitem{TechEdSat}
``Technology educational satellite series: Techedsat-6,'' National Aeronautics
  and Space Administration, Nov 2017.

\bibitem{Audacy}
\BIBentryALTinterwordspacing
(2017) Audacy services. Audacy Coporation. [Online]. Available:
  \url{https://audacy.space}
\BIBentrySTDinterwordspacing

\bibitem{IridiumFCCStatement}
\emph{Iridium NEXT Engineering Statement}, Federal Communications Commission, 8
  2016, appendix to application to modify existing license.

\bibitem{IridiumLatency}
M.~M. McMahon and R.~Rathburn, ``Measuring latency in iridium satellite
  constellation data services,'' \emph{US Naval Academy Report}, no. A291464,
  2005.

\bibitem{GlobalstarFCC}
\emph{Iridium NEXT Engineering Statement}, Federal Communications Commission, 3
  1996, appendix to application to launch and operate.

\end{thebibliography}
%



\end{document}